\documentclass[12pt]{iopart}

\usepackage{iopams}
\usepackage{graphicx}

\begin{document}

\title{Addendum to `Algebraic equations for the exceptional eigenspectrum of the generalized Rabi model'}

\author{Zi-Min Li$^{1}$ and Murray T. Batchelor$^{1,2,3}$}

\address{$^{1}$Centre for Modern Physics, Chongqing University, Chongqing 400044, China}

\address{$^{2}$Department of Theoretical Physics, 
Research School of Physics and Engineering, Australian National University, Canberra, ACT 0200, Australia}

\address{$^{3}$Mathematical Sciences Institute, Australian National University, Canberra ACT 0200, Australia}

\ead{batchelor@cqu.edu.cn}

\begin{abstract}
In our recent paper (Li and Batchelor 2015 {\sl J. Phys. A: Math. Theor.} \textbf{48} 454005) we obtained 
exceptional points in the eigenspectrum of the generalized Rabi model in terms of  
a set of algebraic equations.
We also gave a proof for the number of roots of the constraint polynomials defining these exceptional solutions 
as a function of the system parameters 
and discussed the number of crossing points in the eigenspectrum.
This approach however, only covered a subset of all exceptional points in the eigenspectrum. 
In this addendum, we clarify the distinction between the exceptional parts of the eigenspectrum for this model 
and discuss the subset of exceptional points not determined in our paper.
\end{abstract}

In a recent paper \cite{LB2015}, we considered the generalized quantum Rabi model 
\begin{equation}
H=\omega \, a^{\dagger}a + g \, \sigma_x(a^{\dagger}+a)+ \Delta \, 
\sigma_z+\epsilon\,\sigma_x  \label{ham}
\end{equation}
and derived a set of algebraic equations whose roots define the 
exceptional points in the eigenspectrum via constraint 
polynomials.\footnote{The parameter $\epsilon$ induces conical intersections at crossing points 
in the energy spectrum~\cite{cones}.}
For this model, the exceptional points are defined by points with energy 
$E = N \omega - g^2/\omega + \epsilon$ or $E = N \omega - g^2/\omega - \epsilon$ 
where $N \ge 0$ is an integer \cite{epsilon}. 
The parameter values at the exceptional points satisfy constraint relations.  
It is known for $\epsilon=0$ -- the standard quantum Rabi model -- that the set of all exceptional points can be split into two 
components \cite{MPS,Braak_review}.
We label these components by ${\cal S}_1$ and ${\cal S}_2$.
For subset ${\cal S}_1$ the constraint relations among the system parameters are polynomials.
They can also be defined in terms of a set of algebraic equations.
These are the exceptional points discussed in \cite{LB2015} for $\epsilon \ne 0$. 
However, the subset ${\cal S}_2$ of exceptional points was not discussed in \cite{LB2015}. 
These exceptional points do not appear to be obtainable in terms of algebraic equations.
Their constraint relations are more complicated and have been discussed recently for the 
case $\epsilon=0$ \cite{MPS, Braak_review}.
Here we use the analytic solution obtained in terms of Frobenius series by Braak \cite{Braak}  
for the energy eigenspectrum of the generalized quantum Rabi model  (\ref{ham}) 
to map out the constraint relations for both types of exceptional points.

\begin{figure}[t]
\begin{center}
\includegraphics[width=0.24\columnwidth]{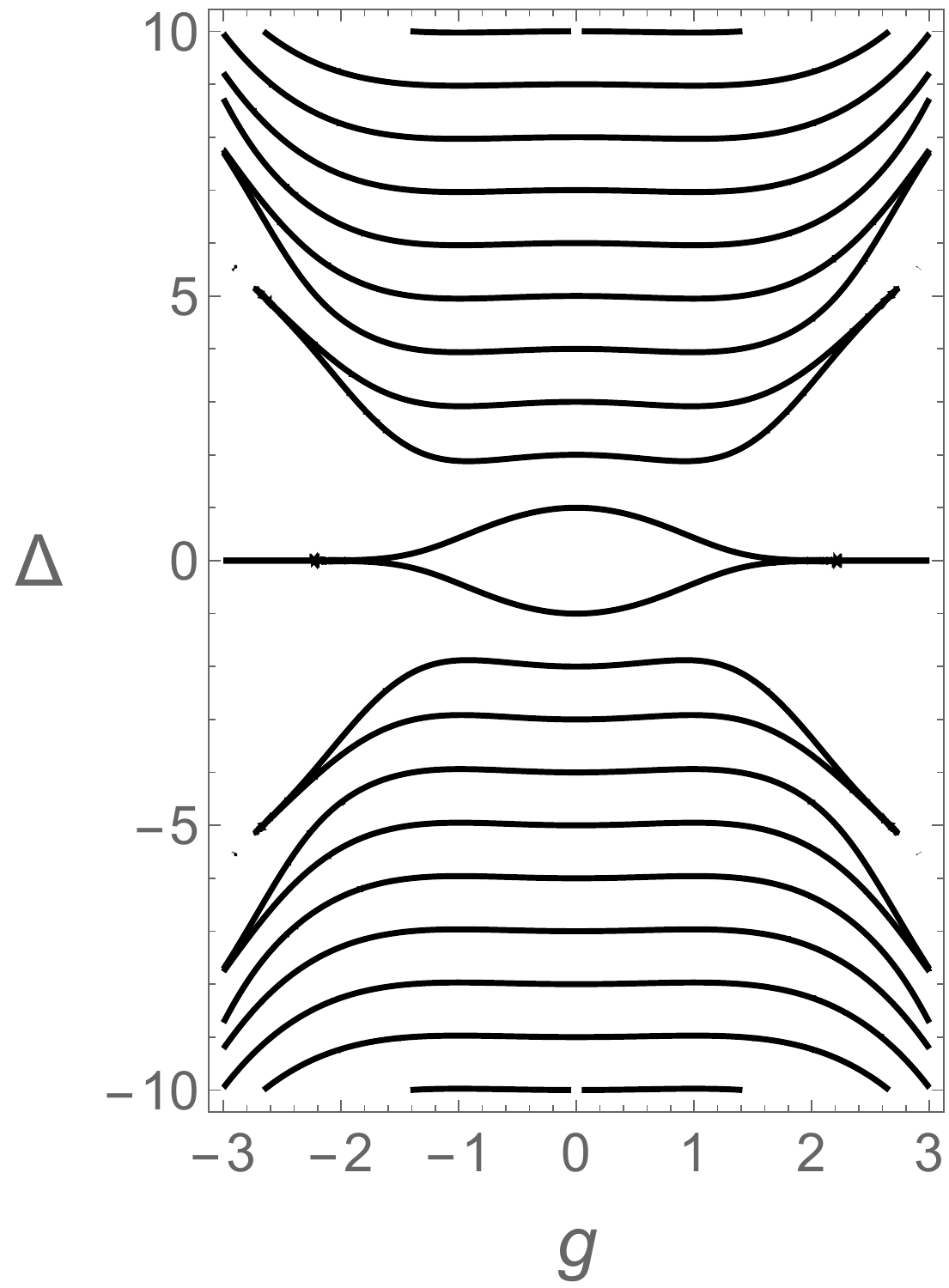}
\includegraphics[width=0.24\columnwidth]{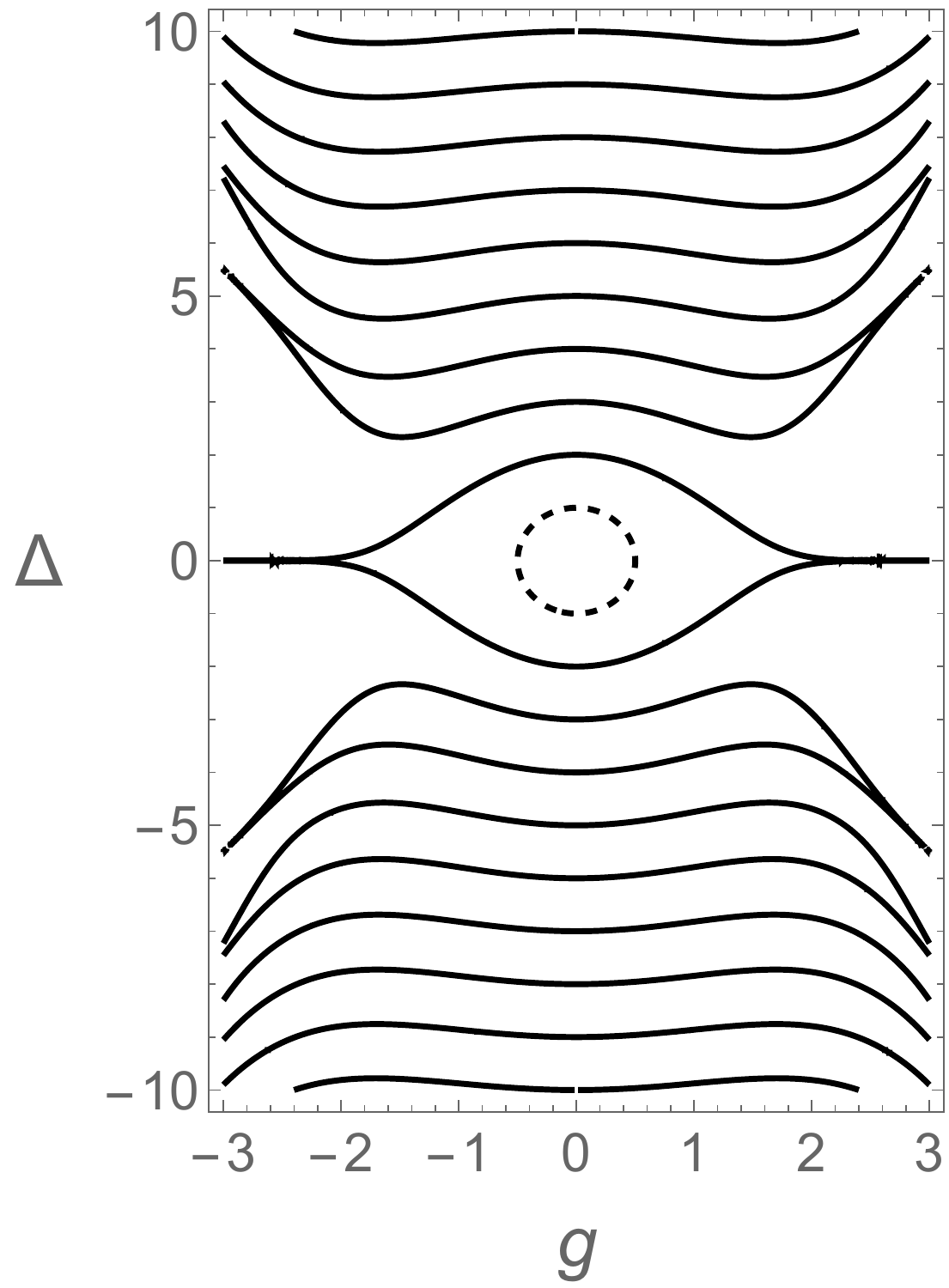} 
\includegraphics[width=0.24\columnwidth]{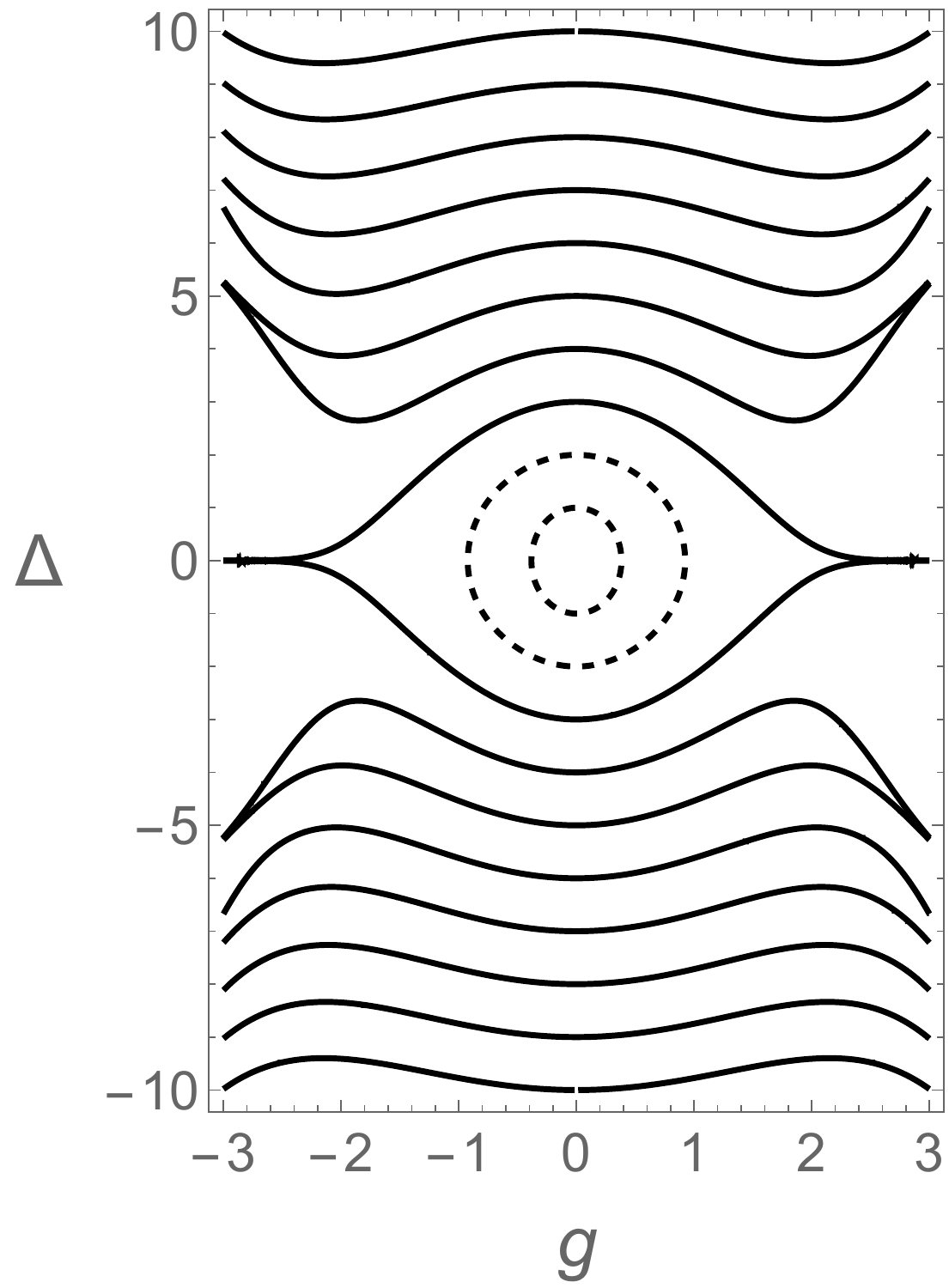}
\includegraphics[width=0.24\columnwidth]{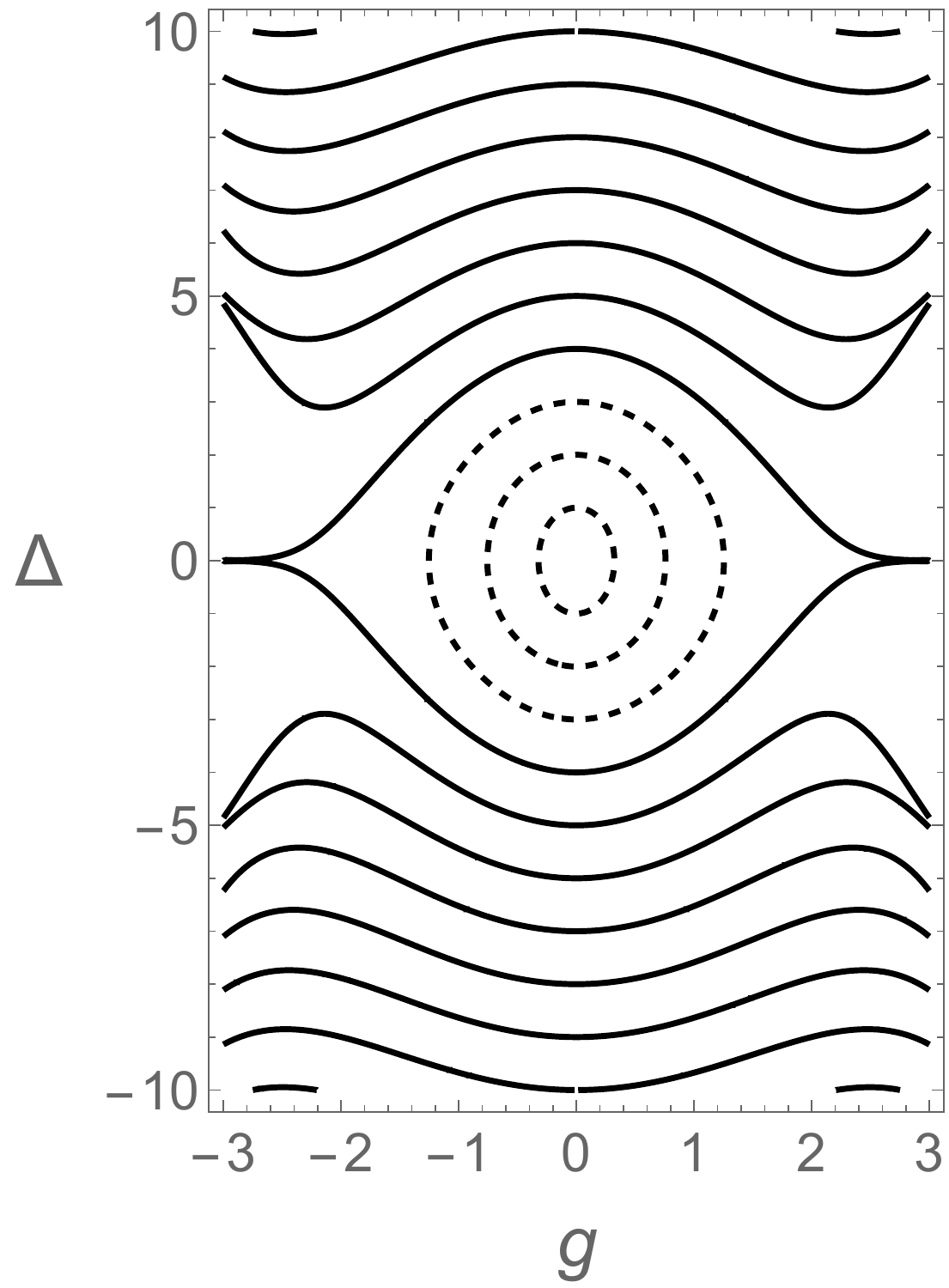}
\caption{Constraint curves for the exceptional points of the quantum Rabi model ($\epsilon=0$) 
in the $\Delta-g$ plane with $\omega=1$. From left to right, the curves have energy $E = N - g^2$ for $N = 0, 1, 2, 3$. 
The closed (dashed) curves correspond to the subset ${\cal S}_1$ of exceptional points 
with the subset ${\cal S}_2$ represented by the infinitely many other (open) curves. 
Subset ${\cal S}_1$ 
corresponds to two-fold degenerate energy eigenvalues. 
}
\label{fig1}
\end{center}
\end{figure}

\begin{figure}[t]
\begin{center}
\includegraphics[width=0.49\columnwidth]{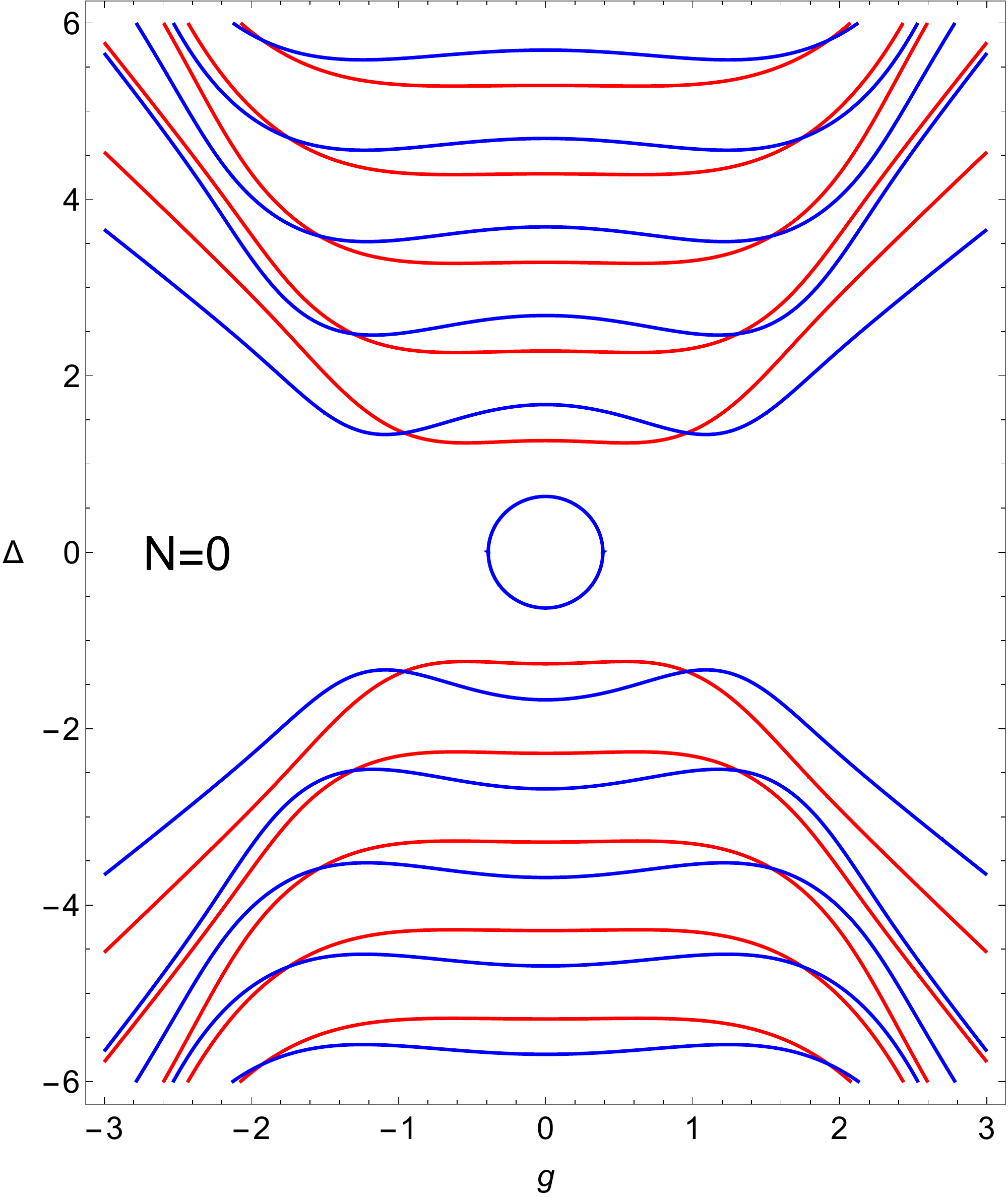}
\includegraphics[width=0.49\columnwidth]{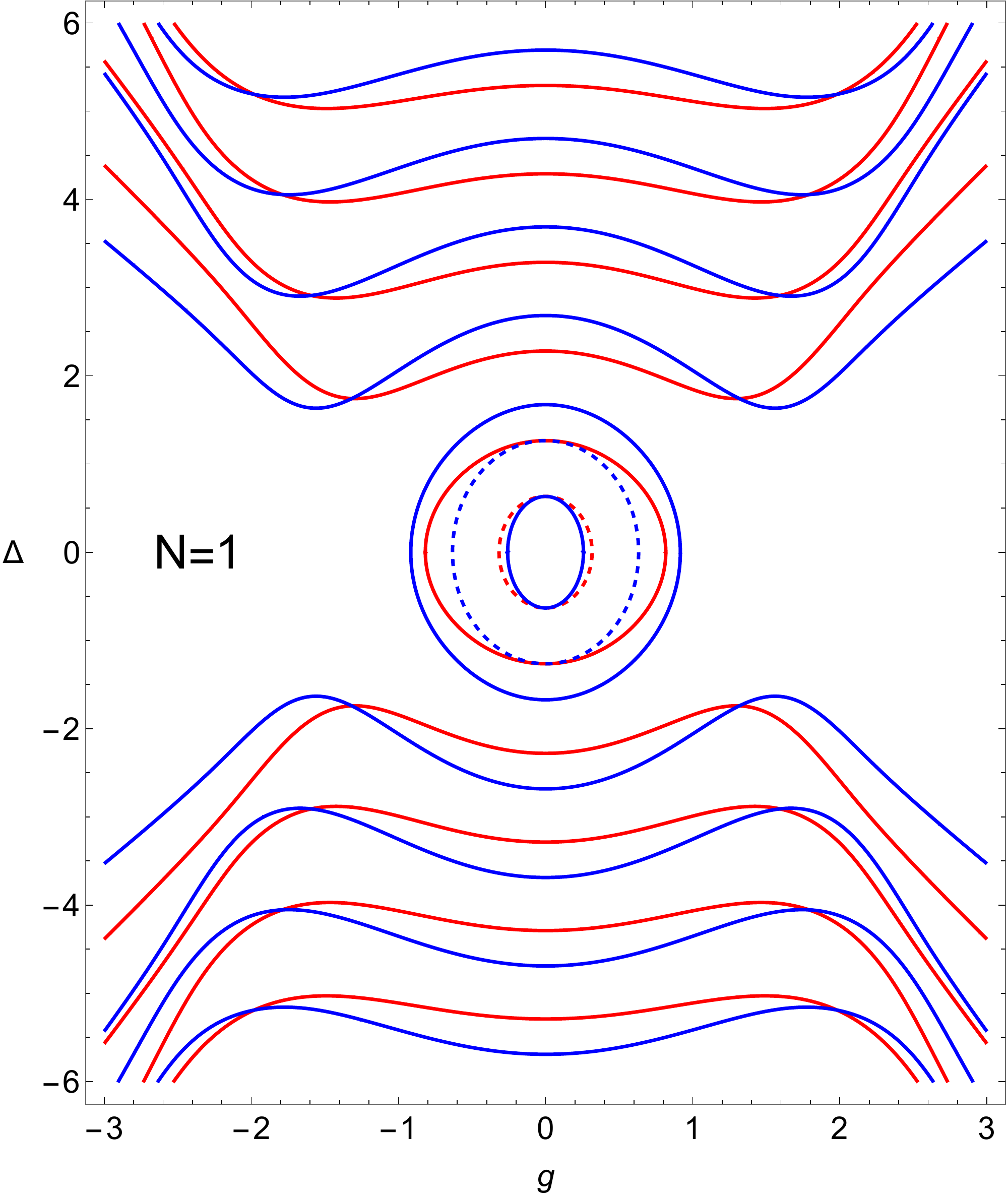}
\vskip 1mm
\includegraphics[width=0.49\columnwidth]{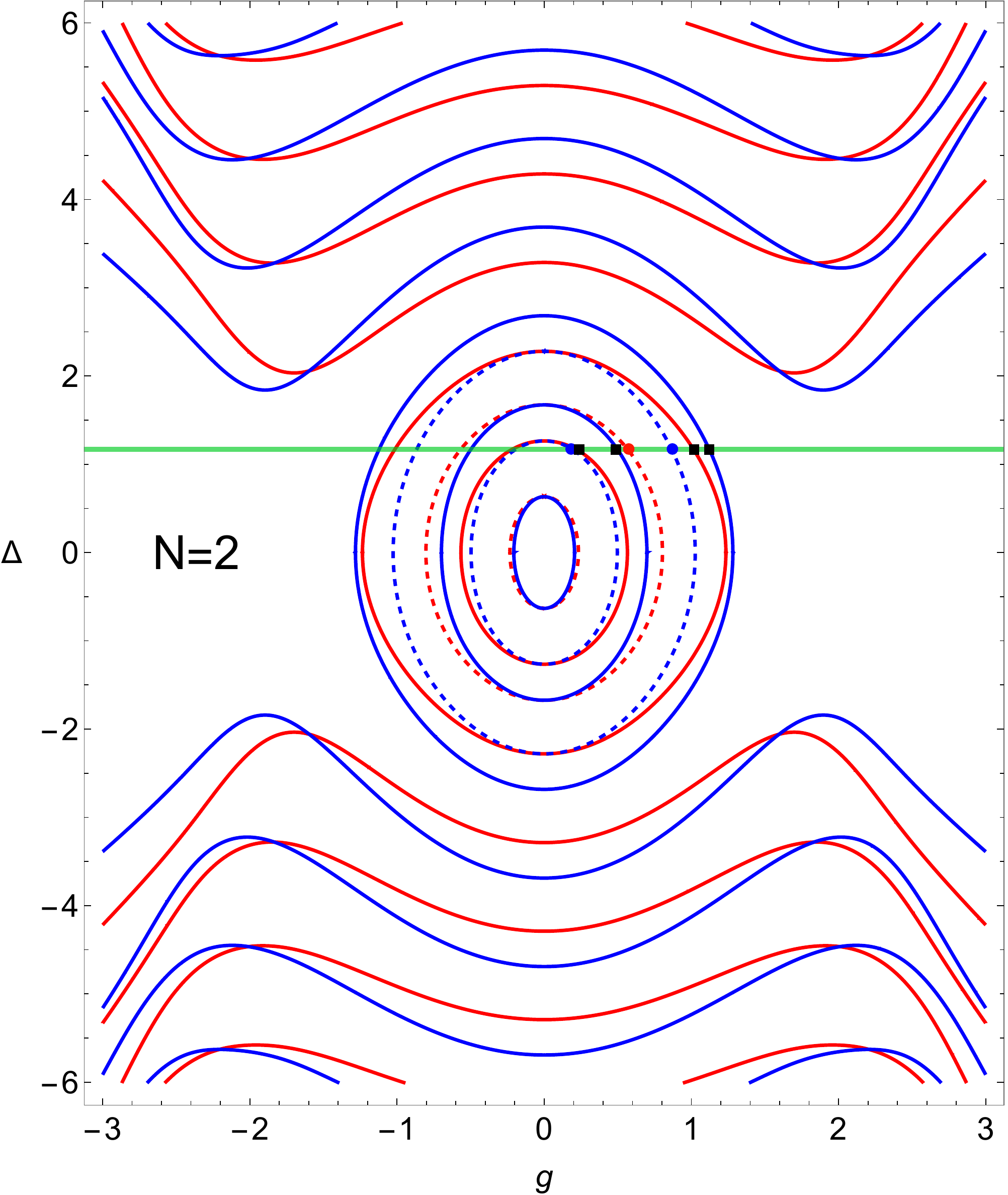}
\includegraphics[width=0.49\columnwidth]{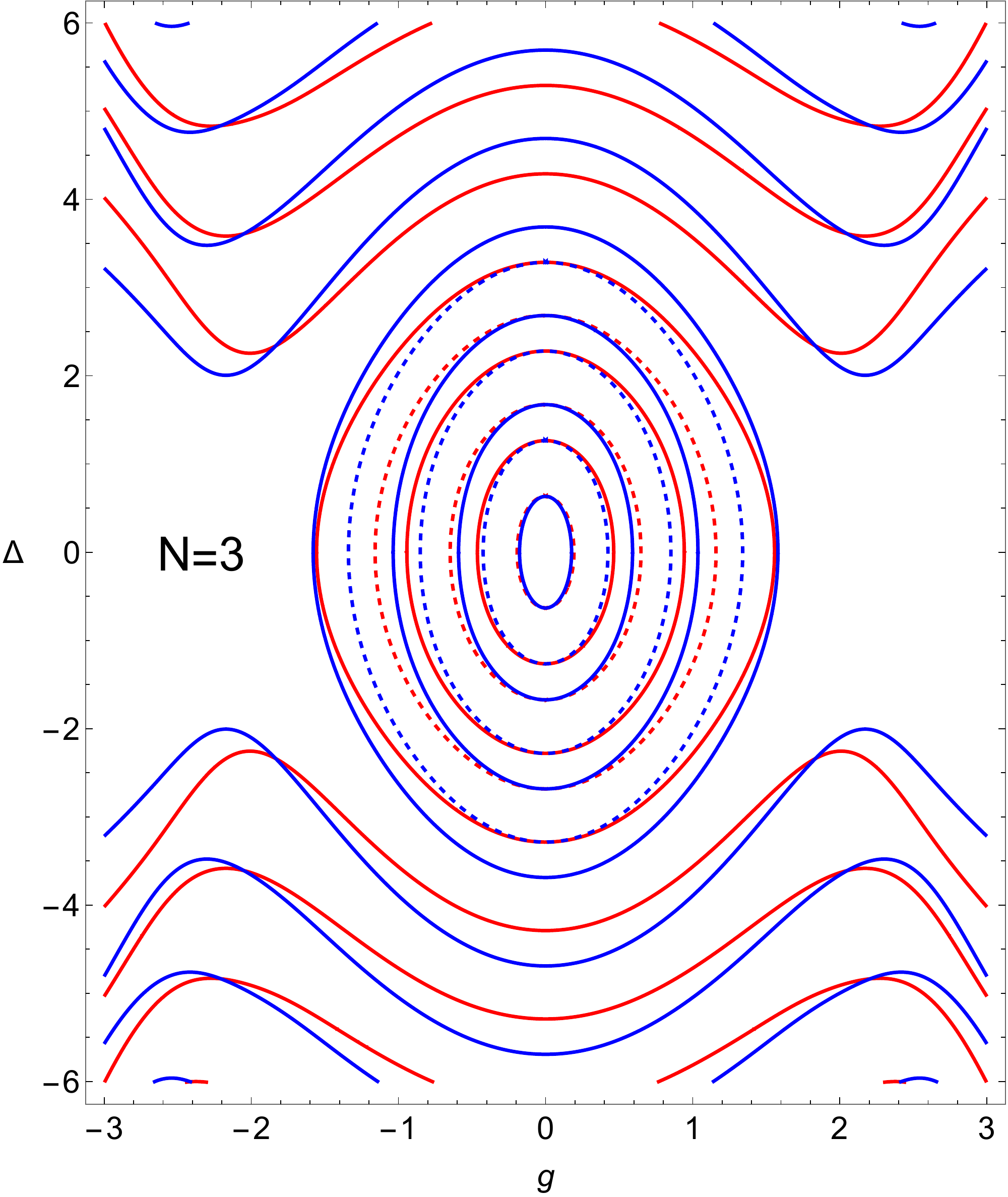}
\caption{Constraint curves for the exceptional points of the generalized quantum Rabi model 
in the $\Delta-g$ plane for parameter values $\omega=1$ and $\epsilon=0.3$. 
Blue curves have energy $E = N - g^2 + \epsilon$ and red curves have energy $E = N - g^2 - \epsilon$. 
As for the $\epsilon=0$ case the closed (dashed) curves correspond to the subset ${\cal S}_1$ of exceptional points 
with the subset ${\cal S}_2$ represented by the other infinitely many curves. 
Subset ${\cal S}_1$ follows from solutions to constraint polynomials.
The seven exceptional points indicated for $N=2$ on the green line $\Delta=1.2$ are readily located in Figure 3. 
}
\label{fig2}
\end{center}
\end{figure}

The $N$th eigenvalue of the hamiltonian (\ref{ham}) 
is given by $E_N = x_N - g^2/\omega$, where $x_N$  is the $N$th zero of \cite{Braak}
\begin{equation}
G_\epsilon(x) = \Delta^2  \bar{R}^+(x) \bar{R}^-(x) - R^+(x) R^-(x) \,, 
\end{equation}
where
\begin{eqnarray}
R^\pm(x)&=& \sum_{n=0}^\infty K_n^\pm(x)  \left( \frac{g}{\omega} \right)^n , \\
\bar{R}^\pm(x)&=& \sum_{n=0}^\infty \frac{K_n^\pm(x)} {x - n \, \omega \pm \epsilon} \left( \frac{g}{\omega} \right)^n .
\end{eqnarray}
$K_n^\pm(x)$ is defined recursively by
$n K_n^\pm = f_{n-1}^\pm(x) \, K_{n-1}^\pm  - K_{n-2}^\pm$, 
with initial conditions $K_0^\pm=1, K_1^\pm(x)=f_0^\pm(x)$ and 
\begin{equation}
f_n^\pm(x)  = \frac{2g}{\omega} + \frac{1}{2g} \left( n \omega - x  \pm \epsilon + \frac{\Delta^2}{x - n\, \omega \pm \epsilon} \right) .
\end{equation}
These equations have also been derived using Bogoliubov operators \cite{Chen}. 
Alternatively the eigenspectrum can be obtained in terms of Wronskians \cite{epsilon,MPS2}.

In terms of the above solution the 
constraint polynomials defining the subset ${\cal S}_1$ of exceptional points considered 
in \cite{LB2015} can be obtained from the condition $K_n^\pm(x) = 0$, which ensures the cancellation of 
poles at the exceptional points in the solution.
For this model 
we can map out the constraint relations defining the exceptional points by redefining the function $G_\epsilon(x)$ 
to cancel out the poles, namely by setting
\begin{equation}
{\cal G}_\epsilon(x) = G_\epsilon(x)  \prod_{n=0}^\infty (x - n \, \omega - \epsilon) (x - n \, \omega + \epsilon). 
\end{equation}
In the numerical procedure for obtaining the energy spectrum the condition $G_\epsilon(x)=0$ 
can then be replaced by the condition ${\cal G}_\epsilon(x)=0$. 
The infinite sums and products are truncated to obtain the desired level of accuracy.
As an example, the constraint curves are shown as a function of the system parameters $\Delta$ and $g$  
in Figure 1  for $\epsilon=0$.
Such plots~\cite{MPS,comment1,comment2} reveal the two distinct classes of curves. 
In this case there is a finite number of closed curves for each value of $N$ 
corresponding to the subset ${\cal S}_1$ of two-fold degenerate exceptional points. 
These exceptional points follow from solutions to the constraint polynomials or alternatively from solutions to the algebraic equations.
The infinitely many other curves correspond to the subset ${\cal S}_2$ of non-degenerate exceptional points.
The situation for $\epsilon \ne 0$ can be seen in Figure 2. 
%
In contrast to the $\epsilon = 0$ case some points in the subset ${\cal S}_2$ of exceptional points 
are located on closed curves.
We have checked that these points do not follow from the constraint polynomials.

\begin{figure}[t]
\begin{center}
\includegraphics[width=0.9\columnwidth]{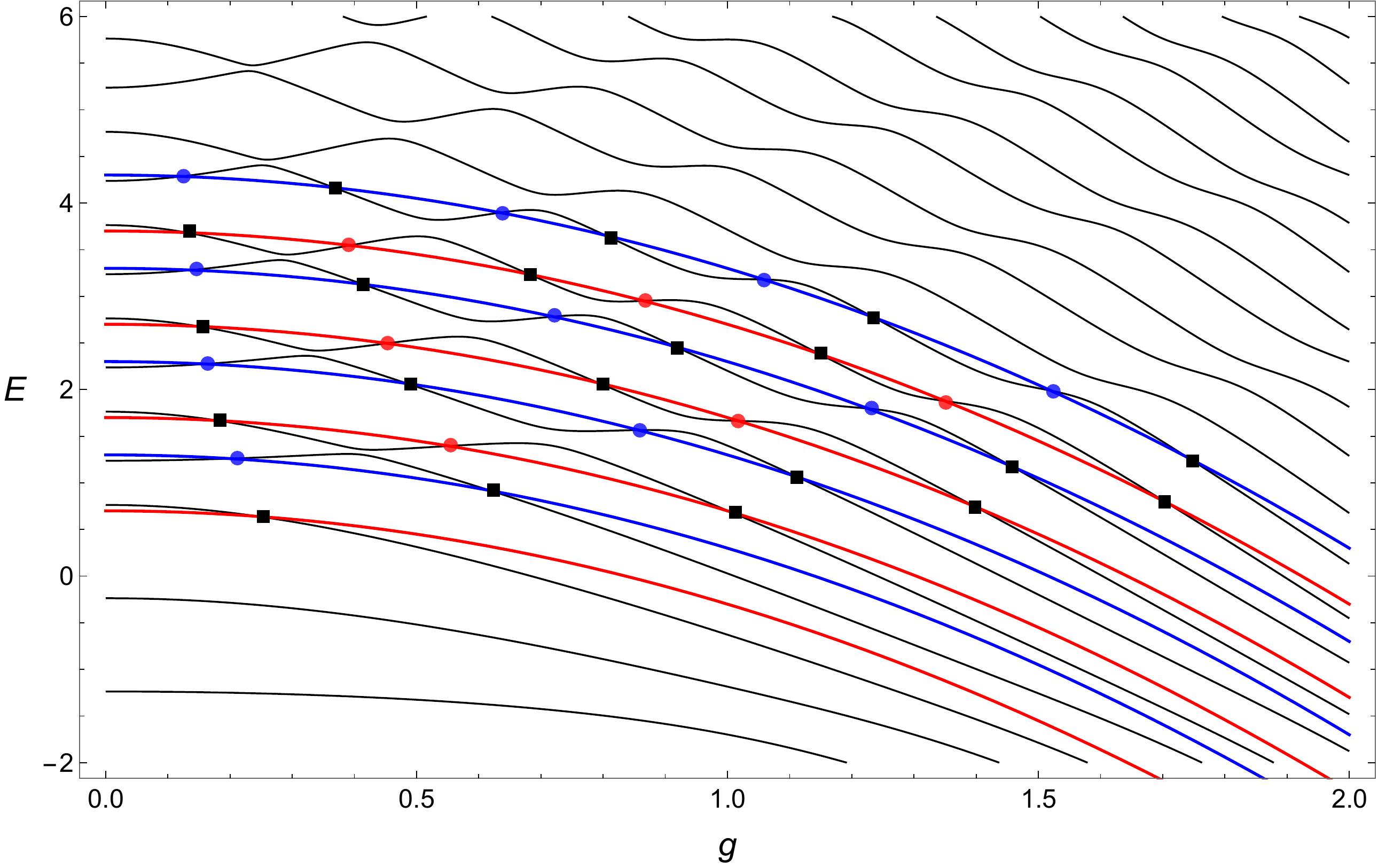}
\caption{The lowest energy levels $E$ in the eigenspectrum of the generalized quantum 
Rabi model as a function of the coupling $g$. 
The parameter values are $\Delta=1.2$, $\omega=1$ and $\epsilon=0.3$. 
The blue lines are the curves $N \omega - g^2/\omega + \epsilon$ for $N=1,\ldots,4$. 
The red lines are the curves $N \omega - g^2/\omega - \epsilon$ for $N=1,\ldots,4$. 
The exceptional points in subset ${\cal S}_1$ are indicated by circles, those in subset ${\cal S}_2$ 
are indicated by squares. 
This figure can be read in conjunction with Figure 2.
}
\label{fig3}
\end{center}
\end{figure}

The corresponding energy spectrum is shown in Figure 3.
Specifically, along the line $\Delta=1.2$ in Figure 2, 
one can read off the corresponding values of $g$ for the exceptional 
points in Figure 3. 
Figure 3 shows an expanded version of Figure 3 in \cite{LB2015}, 
where now {\em all} exceptional points for the lowest few energy levels are indicated. 
In Figure 3 of \cite{LB2015} only exceptional points corresponding to the algebraic solutions or constraint polynomials were shown.
It should be possible to derive a condition in terms of a Wronskian which 
defines the constraint curves for the subset ${\cal S}_2$ of exceptional points, as has been done for 
$\epsilon=0$ \cite{MPS}.
These curves may also be mapped out using the Hill's determinant method~\cite{comment1,Hills} adapted for $\epsilon \ne 0$.
Among the various approaches Braak's solution appears most convenient.
We note that, at the specific parameter values in Figure 3, 
for each value of $N$ there are precisely $N$ exceptional points  in subset ${\cal S}_2$ on each of the curves 
$E = N \omega - g^2/\omega \pm \epsilon$.
However, this is not always true.
For given $\epsilon$ the number of exceptional points varies with $\Delta$, 
as can be seen clearly from Figure 2.

\end{document}